\documentstyle{l-aa}
\begin{document}

\title{The central black hole masses and Doppler factors of the $\gamma$-ray 
loud blazars}

\author{J.H. Fan\inst{1,2} \and G.Z. Xie \inst{3} \and R. Bacon \inst{1}}
\institute{CRAL Observatoire de Lyon, 9 Avenue Charles André, 
 69 563 Saint-Genis-laval Cedex, France
\and Center for Astrophysics, Guangzhou Normal University, 
 Guangzhou 510400, China, e-mail:jhfan@guangztc.edu.cn \and
 Yunnan Observatory, Chinese Academy of Sciences, Kunming 650011, China }

\date{Received<data>;accepted<data>}

\maketitle
\begin{abstract}

 In this paper, The central black hole masses and the Doppler factors are 
 derived for PKS 0528+134, PKS 0537-441, 3C279, PKS 1406-074,  PKS 1622-297,  
 Q1633+382,  Mkn 501, and BL Lacertae. The masses obtained are in the 
 range of (1 - 7)  $\times 10^{7}M_{\odot}$ and compared with that 
 obtained  with the Klein-Nishina cross section considered (Dermer 
 \& Gehrels 1995). If we considered only the Thomson cross section, 
 the masses are in the range of 
 2.6$\times 10^{6}M_{\odot}$ - 2$\times 10^{11}M_{\odot}$.
 The masses obtained from our method are less sensitive to the flux 
 than those obtained from Dermer \& Gehrels (1995) method.  The masses
 obtained from two flares (1991 and 1996 flares) of 3C279 are almost the
 same.  For 3C279 and BL Lacertae,  viewing angle, $\theta$, and 
 Lorentz factor, $\Gamma$, are estimated from the derived Doppler 
 factor and the measured superluminal velocity. For 3C279,  
 $\theta = 10^{\circ}.9-15^{\circ}.6$,
 $\Gamma$ = 2.4-14.4 for $\delta$ = 3.37;  
 $\theta = 8^{\circ}.45-9^{\circ}.7$, $\Gamma$ = 2.95-11.20 for 
 $\delta$ = 4.89; For BL Lacertae,  $\theta = 25^{\circ}-29^{\circ}.4$, 
 $\Gamma$ = 2.0-4.0.

\end{abstract}

\section{Introduction}

 One of the most important results of the CGRO/EGRET instrument in 
 the field of 
 extragalactic astronomy  is the discovery that blazars (i.e., flat-spectrum 
 radio quasars--(FSRQs) and BL Lac objects)  emit most of their bolometric 
 luminosity in the high  $\gamma$-rays ( E $>$ 100 MeV ) energy range.  
 Many of the $\gamma$-ray emitters are also superluminal radio sources 
 (von Montigny et al. 1995).  The common properties of these EGRET-detected 
 AGNs are the  following:  The $\gamma$-ray flux is dominant over the flux 
 in lower energy bands; The $\gamma$-ray luminosity above 100 MeV ranges 
 from less than $3 \times 10^{44}$ erg s$^{-1}$  to more than 
  $10^{49}$ erg s$^{-1}$ 
 (assuming isotropic emission); Many of the sources are strongly variable in 
 the $\gamma$-ray band on timescales from days to months (Mukherjee et al. 
 1997), but large flux  variability on short timescales of $< 1$ day is  
 also detected ( see below ).  Some correlations between $\gamma$-ray and 
 the lower energetic bands are discussed (see Dondi \& Ghisellini 1995; 
 Fan 1997; Fan et al. 1998a;   M$\ddot{u}$cke et al. 1997; Xie et al. 1997; 
 Zhou et al. 1997). These suggest that the $\gamma$-ray emission is  
 likely from the jet. 

 Various of models for $\gamma$-ray emission have been proposed: Namely,
 (1) the inverse Compton process on the external photons ($ ECS$),  in which 
 the soft photons  are directly from a nearby accretion disk ( Dermer et al. 
 1992; Coppi et al. 1993 )  or  from disk radiation reprocessed  in some 
 region of AGNs ( e.g. broad emission line region)  (Sikora et al.  1994; 
 Blandford \& Levinson 1995); 
 (2) the synchrotron self-Compton model ($SSC$), in which the soft photons 
 originate as  synchrotron emission in the jet ( Maraschi et al. 1992; 
 Bloom \& Marscher 1992, 1993;  Zdziarski \& Krolik 1993; Bloom \& Marscher
 1996; Marscher \& Travis 1996); 
 (3) synchrotron emission from ultrarelativistic electrons and positrons 
 produced in a  proton-induced cascade ($PIC$) (Mannheim \& Biermann 1992; 
 Mannheim 1993; Cheng \& Ding 1994). 
 TeV radiations are observed from 3 X-ray-selected BL Lacertae objects (XBLs):
 Mkn 421 ( Punch et al. 1992); Mkn 501 (Quinn et al. 1996), and 
 IES 2344+514 (Catanese et al. 1998). 
 But there is no consensus  yet on the dominant emission process (see 
 $3C273$ for instance, von Montigny et al. 1997); for PKS 0528+134, the lower 
  and higher states can be fitted by different models 
 (eg. B$\ddot{o}$ttcher \& Collmar 1998).
 
 The $\gamma$-rays are produced at a distance of
 $\sim 100 R_{g}$ (Hartman et al. 1996), $205R_{g}$ (Xie et al. 1998), and
 hundreds of Schwarzshild radii (Ghisellini \& Madau, 1996; Celotti 
 \& Ghisellini 1998).  We think that this distance is an important parameter, 
 which can
 be used to constrain the mass of the central black hole. In this paper, 
 we will use it to derive the central black hole mass and the Doppler
 factor for some blazars with short timescales.  The paper is arranged as 
 follows: In section 2, we estimate the mass of the central black hole
 and the Doppler factor; in section 3, we give some discussions and 
 a brief summary.

 H$_{0}$ = 75 km s$^{-1}$ Mpc$^{-1}$, and q$_{0}$ = 0.5 are adopted through
 out the paper.

 \section{The mass of the central black hole and the Doppler factor}

 \subsection{ Data}

 Some $\gamma$-ray loud blazars have been observed several times with EGRET
 while three XBLs have been observed to show TeV radiations. In the following
 section, we present the $\gamma$-ray loud blazars with  available $\gamma$-ray variation
 timescales. Because the variability timescale corresponds to different
 variation amplitude for different source and/or different observation
 period, we use the doubling timescale,  
 $\Delta T_{D} = (F_{initial}/\Delta F) \Delta T$, as the variability
 timescale.

 \subsubsection{PKS 0528+134}

 PKS 0528+134, $z = 2.07$ (Hunter et al. 1993), is one of the most 
 luminous examples of blazars. It is observed by EGRET, COMPTE and OSSE aboard
 the CGRO (see Hunter et al. 1993,  McNaron-Brown et al. 1995, 
 Mukherjee et al. 1996, Collmar et al. 1997; Sambruna et al.  1997).

 During the period of 16-30 May 1991, the source showed 
   F($>$100 MeV) = ( 1.0 $\pm$ 0.2) $\times 10^{-6}$ photon cm$^{-2}$ s$^{-1}$
  with photon spectral index $\alpha_{\gamma} = 2.56\pm0.09$.
 
 During 23-29 March 1993, 
   F($>$100 MeV) = (0.23$\pm$0.12 -- 3.08 $\pm$ 0.35) 
 $\times 10^{-6}$ photon cm$^{-2}$ s$^{-1}$; with a photon spectral 
 index $\alpha_{\gamma}$ = 2.21$\pm$0.10.  In the 1993 observation, 
 a variation of order 100\% over a timescale of $\sim$
 2 days was detected (see Wagner et al. 1997), which suggest a doubling time
 scale of $\Delta T_{D}$ = 1day.

 During  August, 1994, 
 F($>$100 MeV) = (0.32 $\pm$ 0.1) $\times 10^{-6}$ photon cm$^{-2}$ s$^{-1}$; 
 with a photon spectral index $\alpha_{\gamma}$ = 2.70.  

 There is a clear evidence that the spectrum becomes harder when the 
 $\gamma$-ray flux increases.

\subsubsection{PKS 0537-441}
 
 PKS 0537-441, $z$ = 0.896, a candidate of gravitational lens 
 (Surpi et al. 1996 ), is a violently variable object ( Fan \& Lin 1998). 
 The  $\gamma$-ray flux varies from (1.83$\pm$0.91) to 
 (8.98$\pm$1.45) $\times 10^{-6}$ photon cm$^{-2}$ s$^{-1}$ 
 (Mukherjee et al. 1997). A flare of a factor of $\sim$3 
 from 0.35 to 2.0 $\times 10^{-6}$ photon cm$^{-2}$s$^{-1}$ over a time 
 scale of $\sim$ 2 days can be seen from Fig. 3 in  Hartman's paper 
 (Hartman 1996). $\Delta T_{D}$ = 16 hrs.

\subsubsection{1253-055, 3C279}

 3C279 is a well known member  of OVV subclass of blazars.  it is perhaps 
 the prototypical superluminal radio source (Moffet et al. 1972); and the 
 first quasar detected at the  energies of $ > $1 GeV with EGRET/CGRO.  
 The simultaneous variability in  X-rays and $\gamma$-rays ($ > $ 100 MeV) 
 suggest for the first time that they are approximately  cospatial (
 M$^{c}$Hardy 1996). The $\gamma$-ray flux varies from 1.28 to 28.7 
 $\times 10^{-6}$ photon cm$^{-2}$ s$^{-1}$ (Mukherjee et al. 1997). 
 Two $\gamma$-ray flares were detected (see Kniffen et al. 1993; 
 Hartman et al. 1996; M$^{c}$Hardy 1996; Wehrle et al. 1998).

 The 16-28 June 1991 flare showed: 
  F($>$100 MeV) = ( 2.8 $\pm$ 0.4) $\times 10^{-6}$ photon cm$^{-2}$ s$^{-1}$
  with a photon spectral index $\alpha_{\gamma}$ = 1.89$\pm$0.06. A variation 
 of a factor of 4 over 2 days was obtained.

 The January-February 1996 flare showed ( see McHardy 1996; Wehrle et al. 
 1998), 
  F($>$100 MeV) = ( 11.0 $\pm$ 1.) $\times 10^{-6}$ photon cm$^{-2}$ s$^{-1}$
  with a photon spectral index $\alpha_{\gamma}$ = 1.97$\pm$0.07. During this
 flare, a variation 
 of a factor of 4$\sim$5 in a day is observed,  
 $\Delta T$ = 6 hrs (Wehrle et al. 1998).
 
 No obvious spectral index variation has been detected when the flux varied.

\subsubsection{PKS 1406-074}

 PKS 1406-074 has been detected to vary with the $\gamma$-ray flux being in
 the range of 1.54 to 12.76$\times 10^{-6}$ photon cm$^{-2}$ s$^{-1}$
  (Mukherjee et al. 1997). During its $\gamma$-ray flare, a flux of
   F($>$100 MeV) = ( 5.5 $\pm$ 1.4) $\times 10^{-6}$ photon cm$^{-2}$ s$^{-1}$
  with a photon spectral index $\alpha_{\gamma}$ = 2.04$\pm$0.15 and a doubling
 timescale of shorter than 16 hours has been obtained (see Wagner et al. 1995).
 
\subsubsection{PKS 1622-297}
 
 For PKS 1622-297, $z$=0.815, we have very little information in lower 
 energy bands. But it
 is one of the most luminous objects in the $\gamma$-ray region. A peak 
 flux of (17$\pm$3)$\times 10^{-6}$ photon cm$^{-2}$ s$^{-1}$ (E$>$100 MeV) 
 and a
 flux increase by a factor of 2 in 9.7 hours  were observed ( Mattox et al. 
  1997).
    
\subsubsection{Q1633+382, 4C 38.41}
 
 Quasar 1633+382, $z=1.814$, is an LPQ (P$_{opt}=2.6\%$, Moore \& Stockman 
 1984). During 1992 November 17 - December 1 period, it was detected to
 show a flux of 
 F($>$100 MeV) = ( 0.30 $\pm$ 0.06) $\times 10^{-6}$ photon cm$^{-2}$ s$^{-1}$
  with a photon spectral index $\alpha_{\gamma}$ =  1.87$\pm$0.07. 
 The flux varied a factor of 1.5 within 24 hrs, $\Delta T_{D}$ = 16 hrs,
  while the  spectral index did not change. The $\gamma$-ray luminosity is 
 at least  two orders of magnitude larger than the maximum ever observed 
 in any  other band (see Mattox et al. 1993).

\subsubsection{B2 1652+399, Mkn 501} 

 Mkn 501, $z=0.033$, together with other two XBLs
 are three known TeV $\gamma$-ray sources  detected by the Whipple group 
 ( see Quinn et al. 1996; Catanese et al. 1997a; Samuelson et al. 1998; 
 Kataoka et al. 1998).

 During 1995 March-July period, Mkn 501 was observed to show a flux of
  F($>$300 GeV) = ( 8.1 $\pm$ 1.4) $\times 10^{-12}$ photon cm$^{-2}$ s$^{-1}$
  with a photon spectral index $\alpha_{\gamma}$ = 2.2. A variation of
 a factor of 4 over one day is also detected,  $\Delta T_{D}$ = 6 hrs.
 The upper limit corresponds to
 a flux of   F($>$100 MeV) = 1.5 $\times 10^{-7}$ photon cm$^{-2}$ s$^{-1}$
 (see Quinn et al. 1996).  During the 1996 multiwavelength campaign, Mkn 501
 was detected with EGRET a flux of  F($>$100 MeV) =
 $(0.32\pm.13)\times 10^{-6}$ photon cm$^{-2}$ s$^{-1}$ with a photon index
 of 1.6$\pm$0.5 (see Kataoka et al. 1998). During 1997 April 9-19 
 observation, Catanese et al. (1997a) obtained 
 F($>$300 GeV) = ( 40.5 $\pm$ 9.6) $\times 10^{-11}$ photon cm$^{-2}$ s$^{-1}$,
 $\alpha_{\gamma}$ = 2.5, the April 9-15 flux corresponds to
 a flux of  F($>$100 MeV) $<$ 3.6 $\times 10^{-7}$ photon cm$^{-2}$ s$^{-1}$.

 The TeV observations  show that the spectrum softens when the
 source brightens.

\subsubsection{2200+420, BL Lacertae}
 
 2200+420 is the prototype of BL Lacertae class. It is  variable in all
 wavelengths (see Fan et al. 1998b, 1998c; Bloom et al. 1997; B$\ddot{o}$ttcher
 \& Bloom 1998; Madejski et al. 1998). A 14-year period was found in the 
 optical light curve (Fan et al. 1998b). During 1995 January 24 - February 14,
  BL Lacertae showed a flux of 
 F($>$100 MeV) = ( 40 $\pm$12) $\times 10^{-8}$ photon cm$^{-2}$ s$^{-1}$
 with a photon spectral index $\alpha_{\gamma}$ = 2.2$\pm$0.3, the up limit
 flux in higher energy is F($>$300 GeV) $<$ 0.53 $\times 10^{-11}$ 
 photon cm$^{-2}$ s$^{-1}$ ( Catanese et al. 1997b). During 1997 January 
 15/22 observation period, it was detected a flux of  F($>$100 MeV) = 
 ( 171 $\pm$42) $\times 10^{-8}$ photon cm$^{-2}$ s$^{-1}$ with  a photon 
 spectral index $\alpha_{\gamma}$ = 1.68$\pm$0.16 and a dramatic factor 
 of 2.5 increase within a timescale of 8hrs, $\Delta T_{D}$ = 3.2 hrs.  
 Besides, simultaneous optical and $\gamma$-ray flares were
 observed ruling out external scattering models (see Bloom et al. 1997). 

 The observations from the object show that the spectrum of BL Lacertae
 hardens when the $\gamma$-ray flux increases.

\subsection{The central black hole mass and the Doppler factor}

 The objects discussed here show variability time scale of hours to days. 
 The variability could be directly related to shock processes in a jet, 
 far from the accretion disk ( we thank Dr. S. D. Bloom to point out this
  for us). If we take the variability timescale as the measurements of  
 the size, $R$, of the  emission region,  then the $R$ in the jet obeys to 
 the inequality,
\begin{equation}
 R \leq c \Delta T_{D} {\frac { \delta }{(1+z)}} cm
\end{equation}
 where $c$ is the speed of light, $\delta$ the Doppler factor, $z$ the 
 redshift of the source, and $\Delta T_{D}$, in units of second, the 
 doubling time scale. 

 For an object with a mass $M$, the Eddington limit gives (Frank, 
 King \& Raine 1985)
\begin{equation}
 L_{Edd.} \approx 1.26 \times 10^{38}({\frac {M}{M_{ \odot}}}) erg s^{-1}
\end{equation}
 So, we have that the intrinsic luminosity, $L^{in.}$ of a source with a 
 mass of $M$ should satisfy $L^{in}\leq L_{Edd.}$.
  
 In the relativistic beaming frame, the observed luminosity is 
 $L^{ob.}=\delta^{(4+\alpha)}L^{in.}$, $\alpha$ is the energy spectral index,
  which  follows that
\begin{equation}
 L^{ob.} \leq\delta^{(4+\alpha)}L_{Edd.}
 \end{equation}

 Ghisellini \& Madau (1996) obtained that the $\gamma$-rays are emitted 
 within the 
 BLR region, which is  $10^{17-18}$ cm far  from the central source. 
 Hartman et al (1996) obtained that the $\gamma$-rays are produced at 
 a distance  of $\sim$ 100$R_{g}$.
 Recently, Celotti \& Ghisellini  (1998) argued that the $\gamma$-rays 
 are from a region of some hundreds of  Schwarzschild radii from the center. 
 From our previous paper, a distance of 205$R_{g}$ is obtained for the 
 $\gamma$-rays from Mkn 421. 
 In the sense of the theory of accretion (Sunyaev 1975). When $R < 200 R_{g}$,
  the electrons in the accretion flow  become  ultrarelativistic. On the 
 other hand, the mixture of relativistic  electrons and nonrelativistic 
 protons has an adiabatic index  $\gamma < {\frac {5}{3}}$, with such an 
 adiabatic index the transition to  supersonic accretion regime is possible 
 in the region  $R < 200 R_{g}$ (Sunyaev, 1975). So, the 200$R_{g}$ is perhaps
 an important critical point. If we assume that  the $\gamma$-rays are 
 from this place then relations (1), (2),  and (3) give

\begin{equation}
  {\frac {M}{M_{\odot}}}= 5\times 10^{2}{\frac{\delta}{1+z}} \Delta T_{D}
 \end{equation}

\begin{equation}
  L^{ob.} \leq 6.3 \times 10^{40} {\frac { \delta^{(5+\alpha )}}{(1+z)}} \Delta T_{D} erg s^{-1}
 \end{equation}

 It is a common property of the EGRET-detected AGNs to show that their 
 $\gamma$-ray flux is dominant over the flux in  lower energy bands but
 this is not always the case (Mukherjee et al. 1997).  For PKS 0528+134 
 and 3C279, their  $\gamma$-ray luminosity, $L_{\gamma}$, is 0.80$L_{bol.}$ 
 and 0.5$L_{bol.}$, respectively (see Sambruna et al. 1997; Hartman et al. 
 1996). Because we  consider the flare states of the selected objects, 
 we can take the  $\gamma$-ray luminosity to  stand for half of the 
 bolometrical luminosity  approximately, i.e.  
 $L_{\gamma} \sim 0.5 \times L_{bol.}$. So, we have 
\begin{equation} 
 L_{\gamma} \leq 3.15 \times 10^{40} {\frac {\delta^{(5+\alpha )}}{(1+z)}} \Delta T_{D} erg s^{-1}
\end{equation}
 which gives
\begin{equation} 
 \delta \geq [{\frac{L_{\gamma}(1+z)}{3.15\times10^{40} erg s^{-1}\Delta T_{D}}}]^{({\frac{1}{5+\alpha}})}
\end{equation}

 So, from the available $L_{\gamma}$ and $\Delta T_{D}$, we can obtain the
 central black hole mass and the Doppler factor from relations (4) and (7)
 and $\alpha = \alpha_{\gamma} - 1$, $\alpha_{\gamma}$ is the photon spectral
 index.  They are shown in table 1, in which, Col. 1 gives the name;  
 Col. 2, the  redshift; Col. 3. the flux F($>$100MeV) in units of 
 $10^{-6}$ photon cm$^{-2}$ s$^{-1}$,  $\sigma$ is the uncertainty; 
 Col. 4, the photon spectral index,  $\alpha_{\gamma}$ = 2.0
 is adopted for 0537-441 (see Fan et al. 1998a); Col. 5, the doubling time 
 scale in units of hours; Col. 6, $\gamma$-ray luminosity (assuming 
 isotropic emission) in units of  $10^{48}$erg s$^{-1}$; Col. 7, Doppler factor
 estimated from equation (7); Col. 8, the central black hole mass in units of 
 $10^{7}M_{\odot}$; Col. 9, the central black hole mass estimated from 
 Dermer \& Gehrels (1996), thereafter D\&G, in units of $10^{7}M_{\odot}$; 
 Col. 10, the mass 
 estimated directly from Eddington limit in units of  $10^{10}M_{\odot}$

\begin{table*}
\caption{ Mass and Doppler factor for $\gamma$-ray loud blazars}
\begin{tabular}{lccccccccc} 
\hline\noalign{\smallskip}
 $Name$ & $z$ & $ F(\sigma)$ & $\alpha_\gamma$ & $\Delta T_{D}$ & $L_{48}$ &  $\delta$  & $M_{7}$ & $M_{7}^{KN}$ & $M_{10}^{T}$ \\
(1)     & (2)  & (3)  & (4) & (5) & (6) & (7) & (8) & (9) & (10)  \\
\noalign{\smallskip} \hline
0528+134 & 2.07  & 3.08(0.35) & 2.21 & 24.  & 18.4  & 4.96 & 6.97 & 22.98 & 14.6 \\ 
0537-441 & 0.894 & 2.0(0.4)   & 2.0  & 16.  & 3.01  & 3.83 & 5.82 & 3.48  & 2.39\\
1253-055 & 0.537 & 2.8(0.4)   & 2.02 & 12.  & 1.34  & 3.37 & 4.74  & 1.92 & 1.06\\
1253-055 & 0.538 & 11.(1.)    & 1.97 & 6.   & 5.75  & 4.89 & 3.43  & 7.53 & 4.56\\
1406-074 & 1.494 & 5.5(1.6)   & 2.04 & 16.21& 23.7  & 5.57 & 6.52  & 23.6 & 18.81\\
1622-297 & 0.815 & 17.(3.)    & 1.87 & 4.85 & 26.9  & 6.97 & 3.35  & 25.  & 21.35\\
1633+382 & 1.814 & 0.96(0.08) & 1.86 & 16.  & 9.72  & 5.16 & 5.28  & 5.74 & 7.71 \\
1652+399 & 0.033 & $<$0.36    & 2.5  & 6.   & 0.0003& 0.89 & 0.94 & 0.001 & 0.00026\\
2200+420 & 0.07  & 1.71(0.42) & 1.68 & 3.2  & 0.019 & 2.04 & 1.09 & 0.02  & 0.015\\ \hline 
\end{tabular}\\
\end{table*}

 \section{Discussion}

 \subsection{Mass}

 From the high $\gamma$-ray luminosity (assuming isotropic emission) and
 Eddington-limit, one can derive the central black hole mass expression,

 \begin{equation}
 M_{10} \geq {\frac{L_{T}}{1.26\times 10^{48} erg s^{-1}}}
 \end{equation} 
 where, L$_T$ is the bolometric luminosity for emission in the Thomson region,
 M$_{10}$ is the central black hole mass in units of 10$^{10}M_{\odot}$. 
 The derived masses are as high as 10$^{11}M_{\odot}$ for some $\gamma$-ray 
 loud blazars,
 PKS 0528+134, PKS 1406-074, and PKS1622-297 for instance (see Col. 10
 in table 1). 
 However, for high energy $\gamma$-ray emission,  Klein-Nishina effects 
 must be considered. D\&G considered the effect and 
 obtained an expression for the  black hole mass, i.e. their equation (16b),

 \begin{equation}
 M_{8}^{KN} \geq {\frac{3\pi d_{L}^{2}(m_{e}c^{2})}{2 \times 1.26\times 
 10^{46} erg s^{-1}}}{\frac{F(\varepsilon_{l},\varepsilon_{u})}{1+z}}
 ln[2\varepsilon_{l}(1+z)]
 \end{equation} 
 where $F(\varepsilon_{l},\varepsilon_{u})$ is the integrated photon flux 
 in units of 10$^{-6}$ photon cm$^{-2}$ s$^{-1}$ between photon energies  
 $\varepsilon_{l}$ and $\varepsilon_{u}$ in units of 0.511MeV. 
 For the objects considered here,  $M_{7}^{KN}$ is obtained and 
 shown in Col. 9 in table 1.  Table 1 shows that the masses obtained from
 our consideration and those estimated from D\&G method are 
 acceptably similar except for 1622-297 and two low redshift BL Lac
 objects (Mkn 501 and BL Lacertae). For 1622-297 our value is about
 7 times less than that estimated from  D\&G method. If we adopt the
 flux density $2.45 \times 10^{-6}$ photon cm$^{-2}$ s$^{-1}$
  for 1622-297 instead of
 the peak value as did Muhkerjee et al. (1997) and Fan et al. (1998a), then
 the isotropic luminosity is $3.87\times 10^{48}$ erg s$^{-1}$.
 This luminosity suggests that the Doppler factor and mass
 obtained from relations (4) and (7) are respectively 5.01 and 2.41$M_{7}$,
 and the mass estimated from D\&G is then 3.61$M_{7}$. The both
 masses are quite similar in this case.  For Mkn 501 and BL 
 Lacertae, our value is much greater than that estimated from D\&G method.
 For 3C279, our results show that the estimated central black hole masses,
 4.74$M_{7}$ and 3.43 $M_{7}$ for 1991 and 1996 flares respectively, 
 are almost  the same, while the masses estimated from D\&G method are
 1.92$M_{7}$ and 7.53$M_{7}$ for 1991 and 1996 flares  respectively. 
 Table 1 (also see relation (9)) shows that the mass obtained from D\&G
 method is sensitive to the flux, variable flux gives different mass for a 
 source. The mass obtained from our method does not depend on the flux
 so sensitively. For 3C279 1991 and 1996 flares, masses obtained from
 our consideration are almost the same while those obtained from D\&G
 method show a difference of  more or less a factor of 4.  But our method
 depends on the timescales (see relation 4). Since we only considered 
 the objects showing short timescales ( hours) in the present paper, 
 the masses obtained are in a range of (1 $\sim$ 7) $\times 10^{7}M_{\odot}$.

 To fit 3C279 multiwavelength energy spectrum corresponding to 1991
 $\gamma$-ray flare, Hartman employed an accreting black hole of 
 $10^{8}M_{\odot}$, our result of  $4.74\times 10^{7}M_{\odot}$ is similar to
 theirs. 

\subsection{ Beaming factors}

 To explain the extremely high and violently variable luminosity of AGNs, 
 beaming model has been proposed. In this model, the Lorentz factor, $\Gamma$,
  and the viewing angle, $\theta$, are not measurable, but they can be 
 obtained through the measurement of superluminal velocity, $\beta_{app.}$, 
 and the  determination of Doppler factor, $\delta$, which are related with 
 the two unmeasurable parameters, $\Gamma$ and $\theta$, in the forms:
  $\beta_{app}={\frac {\beta_{in} sin \theta}{(1-\beta_{in}cos \theta)} }$,
 $\Gamma={\frac {1}{\sqrt{1-\beta_{in}^{2}}}}$, and 
 $\delta=(\Gamma (1-\beta_{in} cos \theta))^{-1} $. So,
  $\Gamma$ and $\theta$ can be obtained from the following relations: 
 $$\Gamma={\frac {\beta_{app}^{2}+\delta ^{2} +1}{2 \delta}}$$
$$\theta = tan^{-1}({\frac {2 \beta_{app}}{\beta_{app}^{2}+\delta ^{2} - 1}})$$
 From our previous work (Fan et al. 1996), we can get superluminal 
 velocities for
 3C279 and BL Lacertae. When the superluminal velocities and
 the derived Doppler factor are substituted to the above two relations. 
 we found that: For 3C279,  $\Gamma$ = 2.4-14.4 and  
 $\theta  = 10^{\circ}.9 - 15^{\circ}.6$
 for $\delta$ = 3.37; and $\Gamma$ = 2.95-11.20 and  
 $\theta  = 8^{\circ}.45 -  9^{\circ}.7$ for $\delta$ = 4.89. 
 For BL Lacertae $\Gamma$ = 2.- 4. and $\theta  = 25^{\circ} - 29^{\circ}.4$.

 To let the optical depth ($\tau_{\gamma\gamma}$)  be less than unity,
 Doppler factor in $\gamma$-ray region has been obtained for some objects 
 by other authors.
  $\delta \geq 7.6$ for Q1633+382 ( Mattox et al., 1993);  
 $\delta \geq 6.3 - 8.5$
 for 3C279 1996 flare ( Wehrle et al. 1998) and  $\delta \geq 3.9 $ for 
 3C279 1991 flare ( Mattox et al. 1993), $\delta \sim 5 $ is also 
 obtained by Henri et al. (1993); 
 $\delta \geq 6.6 \sim 8.1$ for
 PKS 1622-297 (Mattox et al. 1997).  Our results in table 1 are consistent with
 those results. 

\subsection{Summary}
 In this paper, the central mass and Doppler factor are obtained for 8 
 $\gamma$-ray loud blazars with available short $\gamma$-ray timescales.
 The mass obtained from relation (4) in the present paper is compared with
 that obtained from D\&G method, the masses obtained from two methods
 are similar for 5 out of 8 objects. Our method is not sensitive to the flux
 as much as D\&G method in estimating the central black hole mass. 
 The masses obtained here are in a range of 
 (1 $\sim$ 7) $\times 10^{7}M_{\odot}$, which is from the fact that the time
 scales considered here are in a range of 3.2 to 24 hours.
 For 3C279, the masses obtained from the two flares are almost the same.
  For 3C279 and BL Lacertae, the Lorentz factor and viewing
  angle are estimated.

\acknowledgements{ The authors thank the referee Dr. S.D. Bloom
 for the comments and suggestions. J.H.F. thanks Drs.  B$\ddot{o}$ttcher, Dermer, 
Carter-Lewis, Catanese, Ghisellini, Kataoka, and Wagner for their sending him 
 their publications and information. This work is support by the 
 National Pan Deng Project of  China. }

\cite{}

\end{document}